\documentclass[]{spie}  

\usepackage{float}
\usepackage{multirow}
\usepackage{amsmath,amsfonts,amssymb}
\usepackage{graphicx}
\usepackage[colorlinks=true, allcolors=blue]{hyperref}
\DeclareUnicodeCharacter{202F}{\,}

\title{IFS spectrograph designs for the Wide-field Spectroscopic Telescope: Architecture and performance gains from curved sensors}

\author[a]{C. Cudennec}
\author[a]{A. Jeanneau}
\author[a]{R. Bacon}
\author[b]{T. Lépine}
\author[a]{M. Lehnert}
\affil[a]{Université Claude Bernard Lyon 1, CNRS, Centre de Recherche Astrophysique de Lyon, UMR5574, Saint-Genis-Laval, France}
\affil[b]{Université Jean Monnet, CNRS, Institut d’Optique
Graduate School, Laboratoire Hubert Curien, UMR 5516, Saint-Etienne, France}

\authorinfo{Send correspondence to corentin.cudennec@univ-lyon1.fr}

\begin{document} 
\maketitle

\begin{abstract}
The Wide-field Spectroscopic Telescope\footnote{\url{https://www.wstelescope.com/}} (WST) is a proposed 12-meter segmented facility optimized for seeing-limited observations in the visible and designed to operate both a high-multiplex multi-object spectrograph and a panoramic integral field spectrograph (IFS). The WST IFS concept builds on instruments such as MUSE at the VLT (Very Large Telescope) \cite{MUSE}, using field splitters and image slicers to reformat a large field into pseudo-slits feeding spectrographs with two optimized spectral channels.

This paper presents the spectrograph architecture developed for the WST IFS, aiming to achieve high throughput and image quality over a wide wavelength range in a cost-effective manner. We investigate the use of curved detectors as a means to simplify the spectrograph layout, reduce aberrations, and potentially improve efficiency. This study establishes a promising baseline for the IFS spectrographs and assesses the benefits of incorporating curved sensors that can guide the development of future large-scale integral field spectrographs.
\end{abstract}


\section{INTRODUCTION}

Over the next decade, the imaging capabilities of some of the new facilities (e.g., LSST/VRO, SKAO, CTA, JWST, Euclid, NGRST, Athena) will detect and classify a huge number of astronomical objects. Yet imaging alone is insufficient: unlocking the full scientific potential of these surveys requires spectroscopic follow-up at a scale that no existing facility can provide.\cite{WSTWP} To meet this challenge, a consortium of 23 research institutes and universities across 10 countries is developing the Wide-field Spectroscopic Telescope (WST), a dedicated spectroscopic facility that will be proposed to the European Southern Observatory (ESO) as part of its Expanding Horizons programme for next-generation ground-based infrastructure\footnote{\url{https://next.eso.org/}}.

WST will be equipped with three instruments: two Multi-Object Spectrographs (low and high spectral resolution) and a panoramic Integral Field Spectrograph (IFS) designed to substantially surpass the state of the art across every key figure of merit. Although the WST design builds on other IFS instruments such as MUSE \cite{MUSE_OD} or BlueMUSE \cite{BMUSE_OD}, it differs in its significantly larger étendue and broader spectral range. Spanning 370--930\,nm, it will observe a $3\times3$\,arcmin$^2$ field of view — nine times that of MUSE — while benefiting from an effective collecting area twice as large. Together, these capabilities place the WST IFS in a class of its own among current and planned integral field spectrographs.

\section{WST IFS OVERVIEW}

\subsection{IFS overview}
\label{sec:overview}

\begin{figure} [H]
\begin{center}
\begin{tabular}{c} 
\includegraphics[height=8cm]{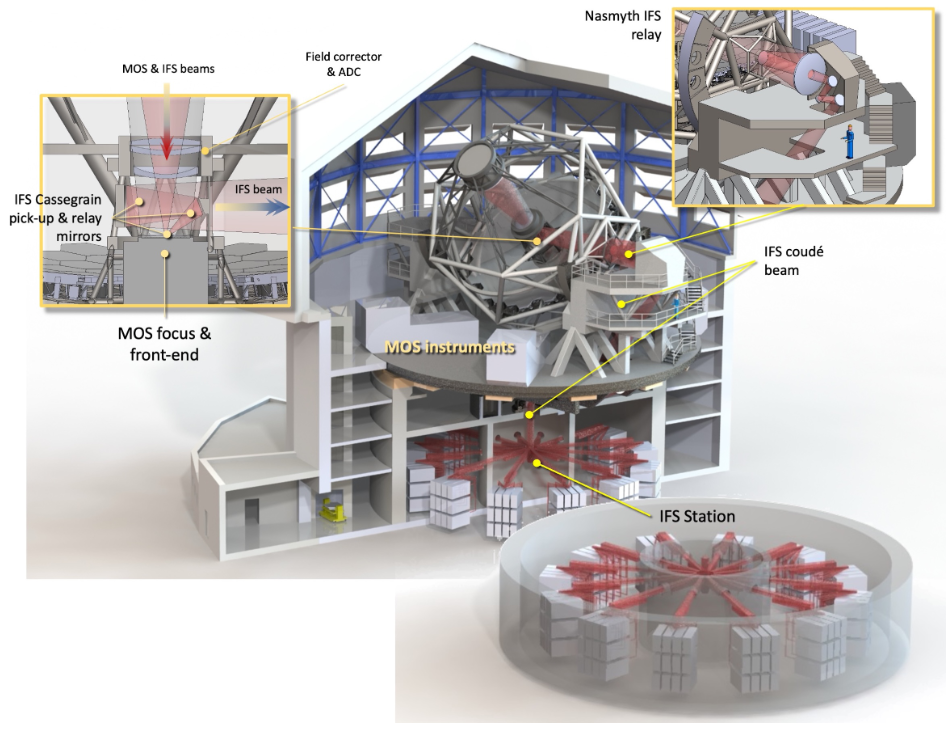}
\end{tabular}
\end{center}
\caption[Facility] 
{ \label{fig:facility} 
Preview of the WST facility (preliminary design).}
\end{figure}

To accommodate the instrument's large volume, the central room at the bottom of the facility is entirely dedicated to the IFS. The light path begins at the Cassegrain focus, passes through one of the Nasmyth platforms, and is relayed to the Coudé focus, located on the telescope's azimuth axis at the center of the IFS room, see Fig.~\ref{fig:facility}.

\begin{figure} [H]
\begin{center}
\begin{tabular}{c} 
\includegraphics[height=6cm]{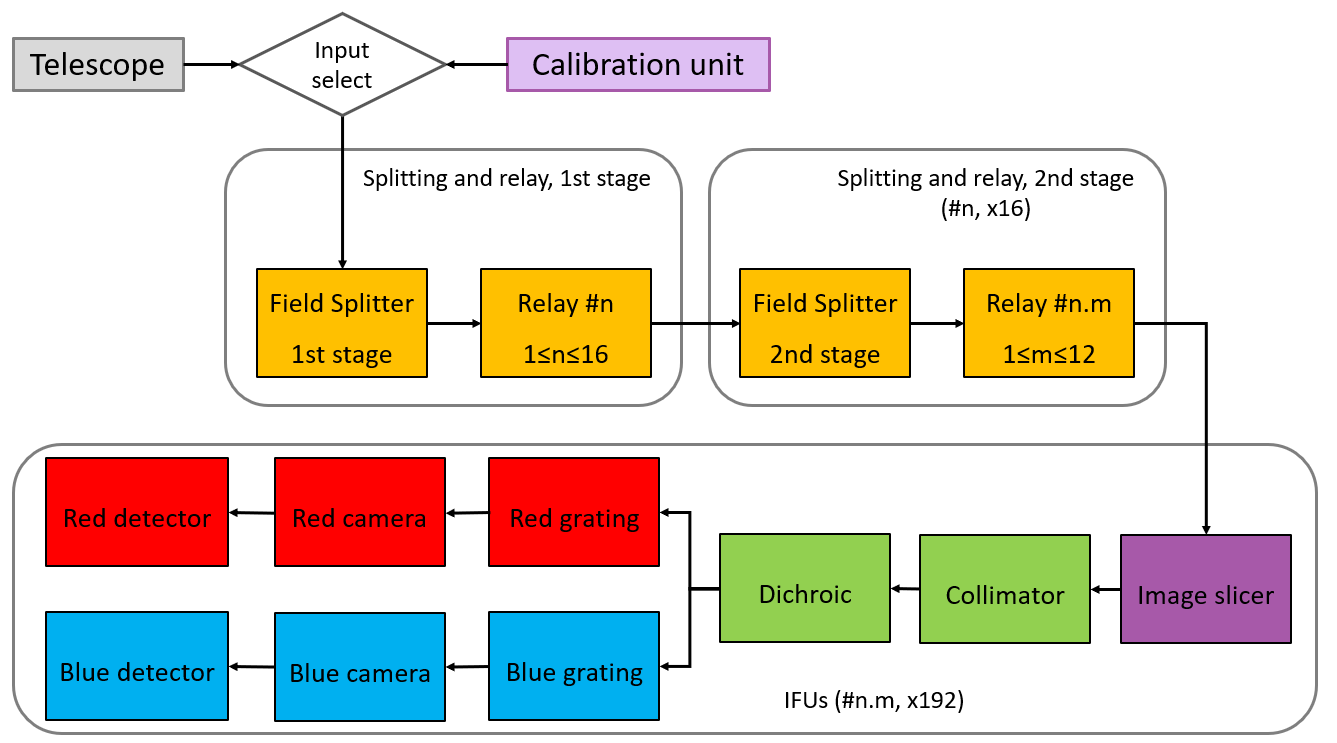}
\end{tabular}
\end{center}
\caption[Scheme] 
{ \label{fig:scheme} 
Overall architecture of the WST integral field spectrograph.}
\end{figure}

From there, the beam is progressively reformatted by a series of subsystems (see Fig.~\ref{fig:scheme}). This integral field spectrograph requires 192 IFUs, which makes a single field-splitting stage impractical, so the field is split into two successive stages. In each resulting IFU, an image slicer dissects the incoming field of view and rearranges it into a long pseudo-slit. This pseudo-slit is then fed into a spectrograph, where a collimator produces a collimated beam that a dichroic splits into two spectral arms: 370–595 nm for the blue arm and 575–930 nm for the red one. Each arm contains a grating, a camera, and a detector.

\subsection{Relay stages}

\begin{figure} [H]
\begin{center}
\begin{tabular}{c} 
\includegraphics[height=5.2cm]{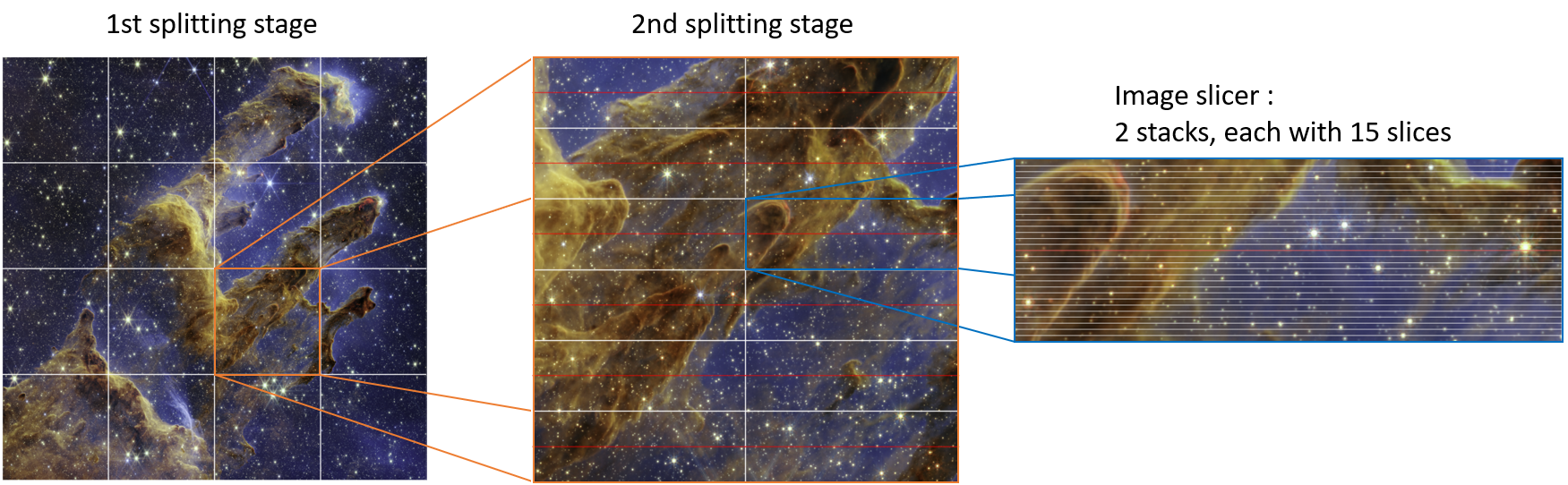}
\end{tabular}
\end{center}
\caption[FSS] 
{ \label{fig:FSS} 
Field-splitting scheme. The science field of view is first divided into a 4$\times$4 grid, after which each sub-field is further split into 2$\times$6 regions. Each of these feeds an image slicer comprising two stacks of 15 slices, with the separation between stacks shown in red. Aspect ratios are preserved.}
\end{figure}

As outlined in Sec.~\ref{sec:overview}, the field must be split into approximately 192 sub-fields before reaching the IFUs. Distributing this operation across two successive splitting stages is both mechanically and optically preferable to attempting it in a single step: concentrating 192 beam paths in one location would create severe accessibility and packaging constraints. Furthermore, this ‘divide-and-conquer’ approach simplifies the design, analysis, manufacturing, and testing of the splitting stage by dividing it into smaller, more manageable units. With two stages, the field is first divided into 16 parts, then each of those into 12 — numbers that are individually smaller than the 24-way split performed by MUSE in a single stage (see Fig.~\ref{fig:FSS}).

\subsubsection*{First splitting stage}

\begin{figure} [H]
\begin{center}
\begin{tabular}{c} 
\includegraphics[height=5.2cm]{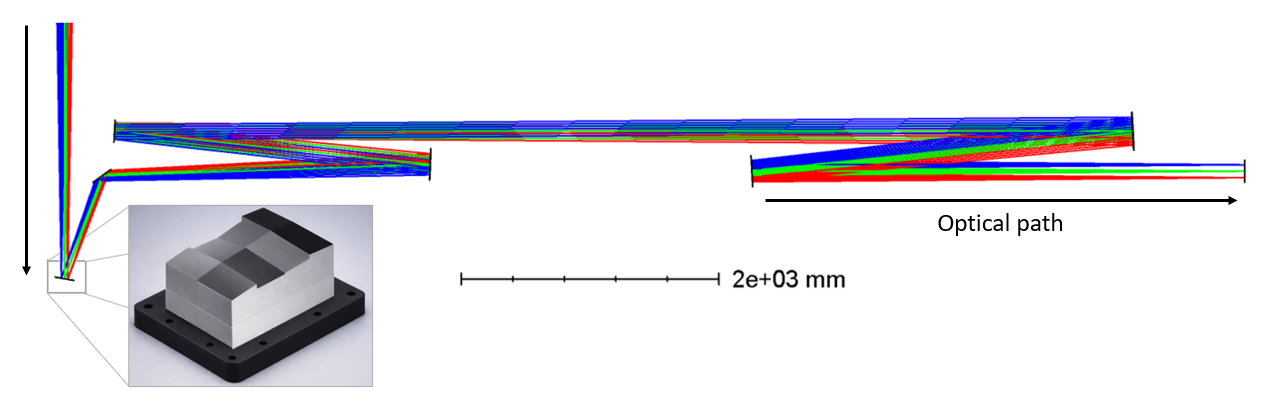}
\end{tabular}
\end{center}
\caption[FSR1] 
{ \label{fig:FSR1} 
The inset shows a field-splitter concept for illustration with only 8 mirrors, rather than the full set of 16.}
\end{figure}

Unlike MUSE, where the beam first encounters a fore-optic before reaching the field splitter, in the WST IFS the beam is split immediately upon arrival. The first splitting stage consists of a single optical component comprising 16 mirrors with individual tilts, an early concept of which is shown in Fig.~\ref{fig:FSR1} with 8 mirrors instead of 16. Each mirror deflects a portion of the beam toward one of 16 relay arms, arranged with radial symmetry about the telescope azimuthal axis (see Fig.~\ref{fig:facility}). The anamorphosis required to correctly sample the spectrograph slits is introduced at this stage, using two toroidal mirrors per relay arm.

\newpage

\subsubsection*{Second splitting stage}

\begin{figure} [H]
\begin{center}
\begin{tabular}{c} 
\includegraphics[height=6.4cm]{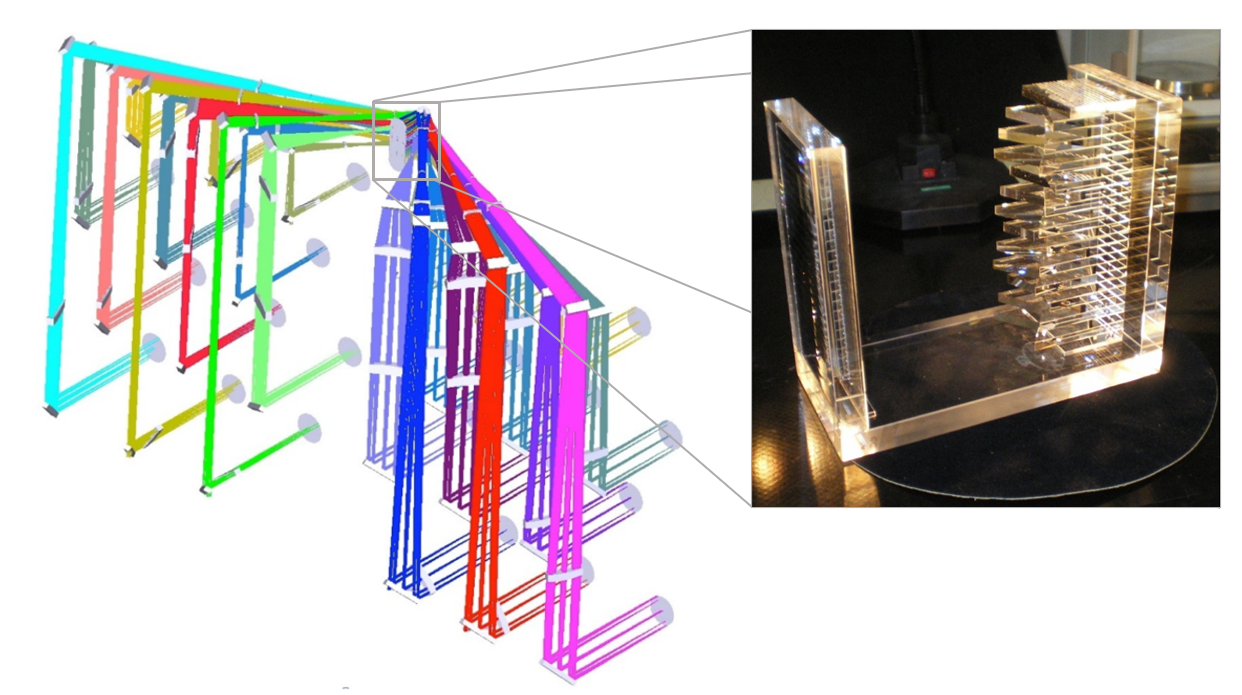}
\end{tabular}
\end{center}
\caption[FSR2] 
{ \label{fig:FSR2} 
Layout of the MUSE splitting stage and relay. WST's second splitting stage will likely have a similar architecture but with a different splitting scheme.}
\end{figure}

The second splitting stage is currently designed along similar lines to its MUSE and BlueMUSE counterparts. As shown in Fig.~\ref{fig:FSR2}, each of the 16 incoming beams encounters a set of 12 lenses and 12 mirrors that divide it further and relay each resulting sub-field to one of the IFUs. The relays currently use fold mirrors and doublets, though replacing these with off-axis mirrors might improve throughput. This stage is therefore still subject to significant evolution, both in optical design and in the physical arrangement of its components.

\subsection{Image slicer}

\begin{figure} [H]
\begin{center}
\begin{tabular}{c} 
\includegraphics[height=6cm]{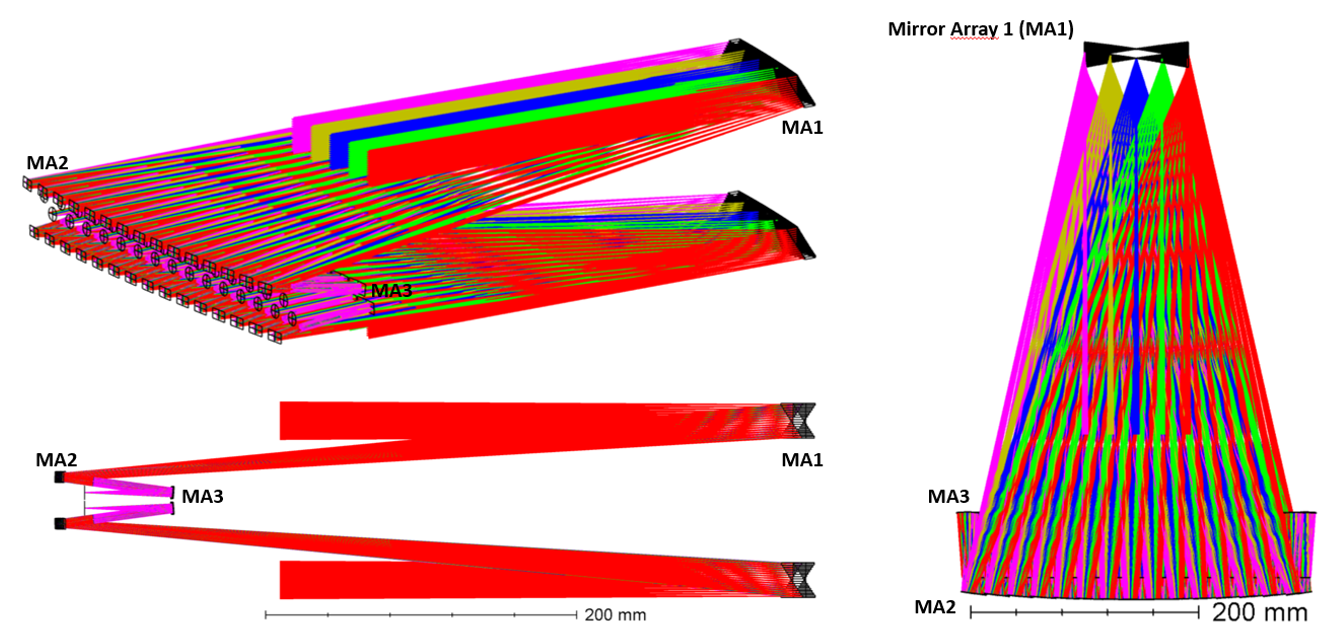}
\end{tabular}
\end{center}
\caption[Slicer] 
{ \label{fig:slicer} 
Preliminary design for the image slicer subsystem.}
\end{figure}

The image-slicer design is still preliminary; however, given the number of spaxels per spectrograph and the camera $f$-number, it is clear that the entrance slit will be long. In most draft designs, the slits are about 300~mm long, or even longer. We therefore chose a three-mirror image slicer, similar to HARMONI’s\cite{Hslicer}, rather than a two-mirror design as in MUSE (see Fig.~\ref{fig:slicer}). The additional row of mirrors (MA2) both provides finer control of the slitlet positions at the slicer output and reduces the incidence angles on the pupil mirrors (MA3). We also chose to split the stack of slices (MA1) into two, again to reduce incidence angles within the slicer, thereby decreasing astigmatism and improving the output image quality.

\section{INITIAL DESIGN FOR A FLAT DETECTOR}

The baseline design with flat detectors uses two spectral arms. A single arm would require the grating to cover the full 370–930 nm range, which would result in low diffraction efficiency across the spectrum. A third arm is unnecessary in terms of spectral resolution and would significantly increase the instrument's cost. We therefore split the spectrum between a blue arm (370-595~nm) and a red arm (575-930~nm), while keeping enough collimated space to place the dichroic and separate the two channels.

\subsubsection*{Collimator design}

A fully dioptric collimator, as in MUSE and BlueMUSE, is not well suited to this geometry for three reasons. First, the collimator would be very long: at f/7 and with a $\sim$200~mm pupil, its length would be about 2800~mm ($2\times7\times200$). Second, we would likely need more than 4 lenses to maintain a long enough collimated path and to balance aberrations between the two arms before the gratings. Third, the optics close to the input slit would become very large; with 3000 spaxels, the slit is already $\sim$300~mm long at f/7.

\begin{figure} [H]
\begin{center}
\begin{tabular}{c} 
\includegraphics[height=5.8cm]{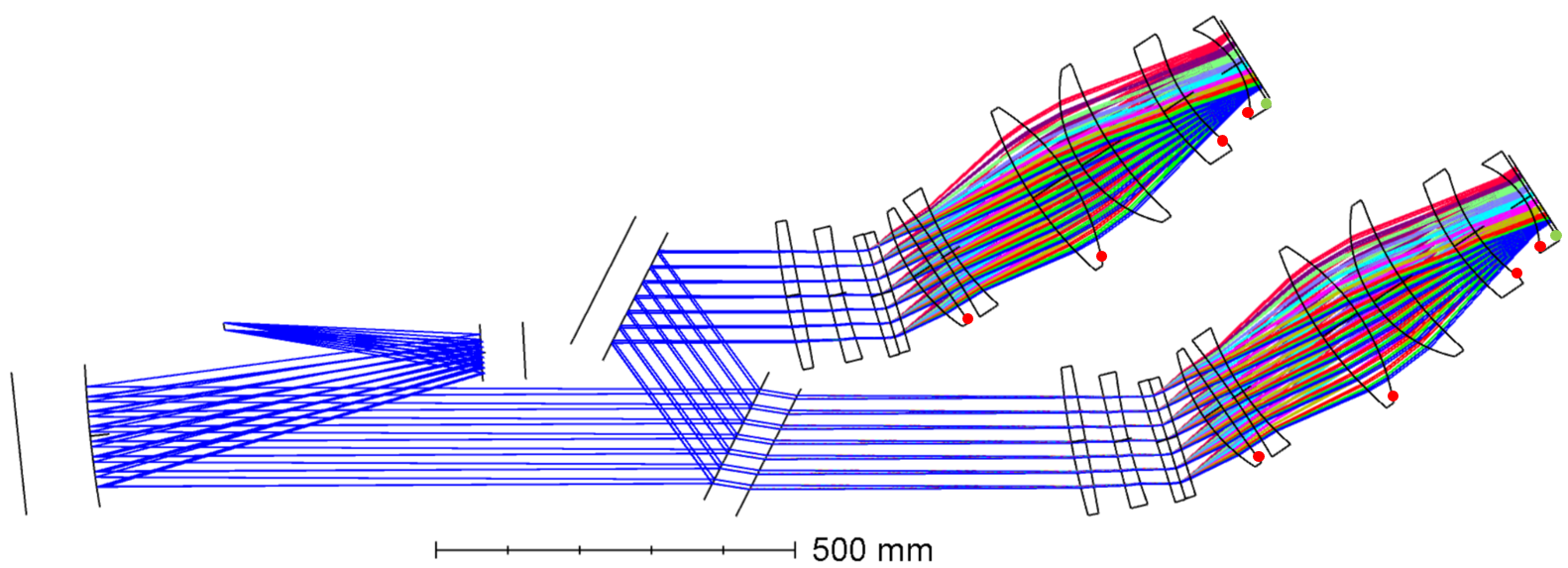}
\end{tabular}
\end{center}
\caption[SF] 
{ \label{fig:SF} 
Spectrograph layout for the design using flat detectors. The red and green dots respectively correspond to aspherics and to cylindrical curvature.}
\end{figure}

For this reason, we adopted a mirror-based collimator inspired by MOS layouts\cite{4MOST} (Fig.~\ref{fig:SF}). Despite the apparent complexity, the collimator uses only two mirrors: a spherical first mirror and a conic second mirror. These elements are the largest of the spectrographs (the second mirror is about 650~mm $\times$ 200~mm). After collimation, a dichroic splits the beam into blue and red channels. Each arm then uses a low-power air-spaced silica doublet. These two lenses have spherical surfaces without wedges or freeform terms. However, they are mounted with small tilts and decentrations with respect to the optical axis to partially compensate for the nodal aberrations --- mostly astigmatism --- generated by the two previous mirrors. This configuration makes the alignment of the collimator insensitive to the optics’ clocking and, interestingly,  requires fewer optical surfaces than a fully dioptric collimator.

As shown in Fig.~\ref{fig:SF}, the input slit has a two-level geometry inherited from the image slicer, whose output slitlets are also distributed over two levels (Fig.~\ref{fig:slicer}). We partially compensate this staircase effect by tilting the input slit, which limits the image quality degradation relative to an ideal thin slit.

\subsubsection*{Camera design}

The cameras for WST must handle a very large geometric \'etendue, driven by the field of view, the primary mirror diameter, and the spaxel size. With pixel pitches of 10\,\textmu m or 15\,\textmu m and the possibility of binning, the slowest camera envisaged operates at f/2.06 and the fastest at f/0.69. The design shown in Fig.~\ref{fig:SF} is based on a 6k--15\,\textmu m detector with $2{\times}1$ binning along the spatial axis, giving focal ratios of f/2.06 and f/2.47 along the spatial and spectral axes respectively. This difference reflects the anamorphism introduced to correctly sample the input slit: the slit is 2.4 pixels wide, and a spaxel is twice the size of a pixel due to binning, yielding an anamorphism ratio of 1.2.

In terms of optical prescription, the first, third, fifth, and sixth lenses each carry one aspheric surface, with departures from the best-fit sphere below 0.4\,mm in all cases. The last lens has a cylindrical second surface, which improves image quality by accommodating the difference in focal ratio between the spatial and spectral axes. This last lens also serves as the cryostat window for the detector and is tilted together with the detector to compensate for axial chromatism.

\subsubsection*{Performance}

\begin{figure} [H]
\begin{center}
\begin{tabular}{c} 
\includegraphics[height=6.5cm]{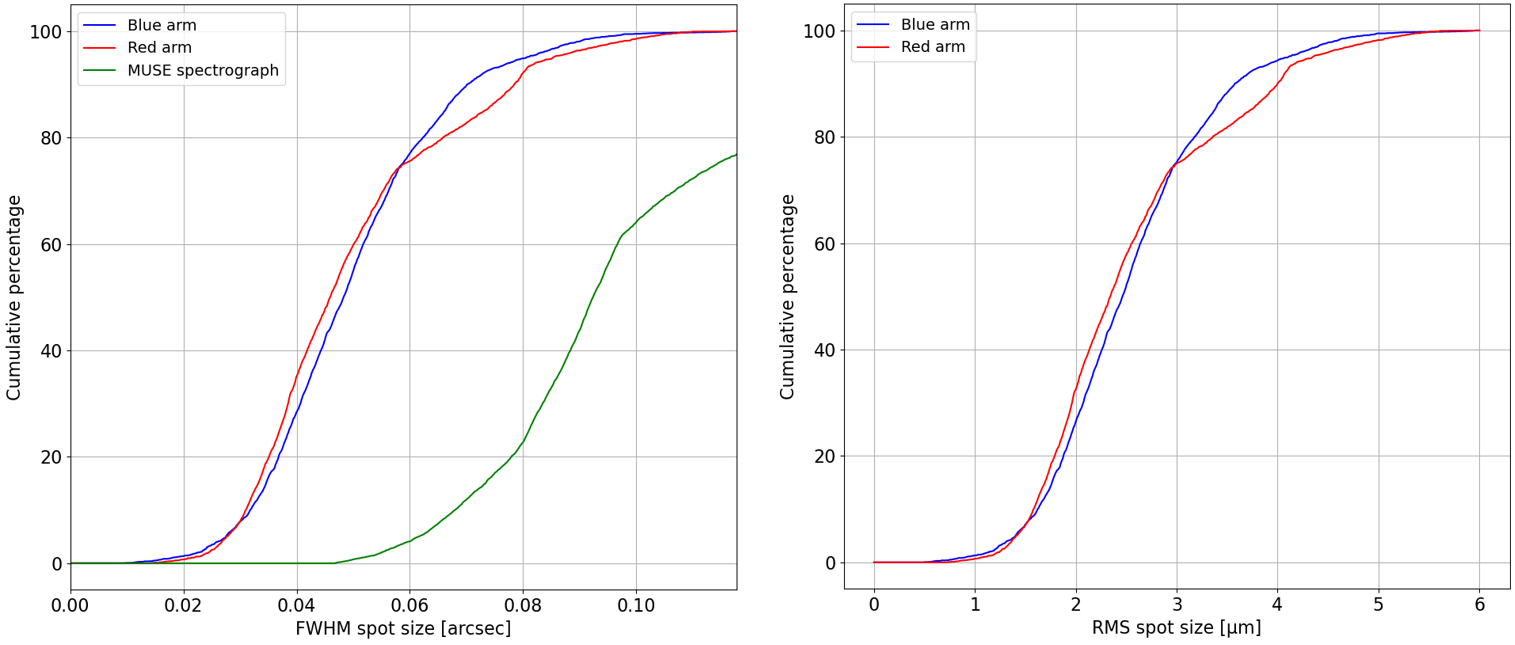}
\end{tabular}
\end{center}
\caption[SF] 
{ \label{fig:IQ_flat} 
Cumulative distribution of the FWHM spot size in the blue and red arms for the \textbf{as-designed (nominal)} spectrograph, projected on sky (left) and on the detector (right).}
\end{figure}

The design delivers excellent image quality, even compared to other IFS spectrographs. In terms of \textbf{as-designed} image quality, the median spot size projected on sky is about 0.048 and 0.046 arcsec FWHM in the blue and the red arms respectively (see Fig.~\ref{fig:IQ_flat}), when MUSE is about 0.092 arcsec FWHM. Although no tolerancing has been performed yet, the relatively modest refractions in the camera suggest that the alignment tolerances will be relatively relaxed.\\

In terms of throughput, the glass selection was driven by the need to minimize internal absorption, particularly at the blue end of the spectrum. Hence, in order to limit the variety of glass types and maximize the throughput, only three materials are used: Silica, CaF$_2$, and PBL35Y. The first two can be sourced from multiple suppliers, while the PBL35Y comes from Ohara with an equivalent available in Schott’s catalog under reference LF5HTi. For the coatings we assume for every optical surface a transmission/reflection coefficient of 99.3\% per optical surface, which is achievable using multidielectric coatings. Thus, excluding the dichroic, the grating diffraction efficiency and the detector quantum efficiency, the design provides a transmission of 86.5\% in the deep blue, 86.7\% above 380 nm in the blue arm, and 86.1\% everywhere in the red arm.

\section{OPTICAL DESIGN WITH CURVED DETECTORS}

\subsection{Field curvature in a spectrograph}
\label{sec:theo}

Curved detectors offer a key advantage: by removing the field curvature constraint during optimization, they free up degrees of freedom that can be used to correct non-stigmatic aberrations.\cite{HighlyCurved,Olaf} However, field curvature in a spectrograph is more complex than in a conventional or off-axis imaging system, because the detector surface must simultaneously accommodate to additional contributions intrinsic to spectrographs.

\subsubsection*{Decomposition of the field curvature}

Along the \textbf{spatial axis} ($x$), the field curvature is simply the sum of two medial 
curvature terms: one from the collimator, $W^{\rm col}_{220m}$, and one from the camera, 
$W^{\rm cam}_{220m}$.

Along the \textbf{spectral axis} ($y$), in addition to the camera's, axial chromatism $\delta_\lambda W_{020}(\lambda)$ also contributes to the field curvature. In fact, since wavelength is mapped to position along $y$ by the grating, this chromatic term effectively bends the focal surface along the spectral direction.

Furthermore, whenever optical elements are used off-axis, they introduce nodal field-curvature contributions. In the general nodal aberration framework \cite{Nodal,Liu}, the medial field curvature wave aberration for a misaligned system takes the form:
\begin{equation}
W(\vec{H}, \vec{\rho}) = W_{220m}\left(\vec{H}_{220m} \cdot \vec{H}_{220m} + b_{220m} \right)(\vec{\rho}\cdot\vec{\rho}) 
= W_{220m}\Big((\vec H-\vec a_{220m})\!\cdot\!(\vec H-\vec a_{220m})+b_{220m}\Big)(\vec\rho\!\cdot\!\vec\rho),
\label{eq:nodal}
\end{equation}
where $\vec{a}_{220m}$ and $b_{220m}$ are normalized coefficients encoding the cumulative effect of all tilts and decentrations $\vec{\sigma}_j$ of each optical element $j$:
\begin{equation*}
\vec{a}_{220m} = \frac{1}{W_{220m}}\sum_j W_{220m,j}\,\vec{\sigma}_j, 
\qquad 
b_{220m} = \frac{1}{W_{220m}}\sum_j W_{220m,j}\,(\vec{\sigma}_j\cdot\vec{\sigma}_j) 
- \vec{a}_{220m}\cdot\vec{a}_{220m},
\end{equation*}
These terms shift the center of curvature away from the optical axis, introducing both a tilt and an asymmetric deformation into the ideal detector shape. Accounting for all contributions : collimator, camera, and axial chromatism, within fourth-order aberration theory, the focus condition that the image surface must satisfy is:
\begin{equation}
0 = \left[W_{020} + \delta_\lambda W_{020}(\lambda) 
+ W^{\rm col}_{220m}\!\left(\vec{H}^{\rm col}_{220m}\cdot\vec{H}^{\rm col}_{220m} 
+ b^{\rm col}_{220m}\right) 
+ W^{\rm cam}_{220m}\!\left(\vec{H}^{\rm cam}_{220m}\cdot\vec{H}^{\rm cam}_{220m} 
+ b^{\rm cam}_{220m}\right)\right](\vec{\rho}\cdot\vec{\rho}),
\label{eq:focus_condition}
\end{equation}
Applying $\vec{\nabla}_{\vec{\rho}}$ to Eq.~\ref{eq:focus_condition} yields the corresponding expression for transverse aberrations:
\begin{equation}
-W_{020}\,\vec{\rho} = \left[\delta_\lambda W_{020}(\lambda) 
+ W^{\rm col}_{220m}\!\left(\vec{H}^{\rm col}_{220m}\cdot\vec{H}^{\rm col}_{220m} 
+ b^{\rm col}_{220m}\right) 
+ W^{\rm cam}_{220m}\!\left(\vec{H}^{\rm cam}_{220m}\cdot\vec{H}^{\rm cam}_{220m} 
+ b^{\rm cam}_{220m}\right)\right]\vec{\rho},
\label{eq:transverse}
\end{equation}
Since the spectrograph is plane-symmetric with respect to the $YOZ$ plane, all tilts and decentrations occur only around the $X$ axis and/or along the $Y$ axis. Consequently, $\vec{a}^{\rm col}_{220m}$ and $\vec{a}^{\rm cam}_{220m}$ have components only along $\vec{y}$.\\
Using $W_{020} = -\frac{\mathrm{NA}^2}{2n}\Delta z$, the detector sag can be written explicitly as:
\begin{equation}
\Delta z = \frac{2n}{\mathrm{NA}^2}\left[
\delta_\lambda W_{020}(\lambda) 
+ W^{\rm col}_{220m}\!\left(x^2 + \left(0 - a^{\rm col}_{220m}\right)^2 + b^{\rm col}_{220m}\right) 
+ W^{\rm cam}_{220m}\!\left(x^2 + \left(y - a^{\rm cam}_{220m}\right)^2 
+ b^{\rm cam}_{220m}\right)
\right],
\label{eq:sag}
\end{equation}
where $\Delta z$ is the image surface sag, and $n$ and $\mathrm{NA}$ are respectively the refractive index and numerical aperture in image space.

\subsubsection*{Mapping wavelength to detector position}

To express $\Delta z$ as an explicit function of detector coordinates alone, we need to convert wavelength $\lambda$ into the spectral position $y$. A second-order Taylor expansion of the axial chromatism $\delta_\lambda W_{020}$ around $\lambda_0$ gives:
\begin{align}
\Delta z = \frac{2n}{\mathrm{NA}^2}\Bigg[
&\frac{d(\delta_\lambda W_{020})}{d\lambda}\bigg|_{\lambda_0}(\lambda-\lambda_0) 
+ \frac{1}{2}\frac{d^2(\delta_\lambda W_{020})}{d\lambda^2}\bigg|_{\lambda_0}(\lambda-\lambda_0)^2 
+ o(\lambda^2) \nonumber\\
&+ W^{\rm cam}_{220m}\left(y - a^{\rm cam}_{220m}\right)^2 
+ \left(W^{\rm col}_{220m} + W^{\rm cam}_{220m}\right)x^2 \nonumber\\
&+ W^{\rm col}_{220m}\!\left({a^{\rm col}_{220m}}^{\!2} + b^{\rm col}_{220m}\right) 
+ W^{\rm cam}_{220m}b^{\rm cam}_{220m}
\Bigg],
\label{eq:sag_taylor}
\end{align}
Using the grating equation, a second-order expansion of $(\lambda - \lambda_0)$ in $y$ yields:
\begin{equation}
\lambda - \lambda_0 = \frac{\cos\beta_0}{pf}\frac{D}{2}\,y 
\;-\; \frac{\sin\beta_0}{2pf^2}\!\left(\frac{D}{2}\right)^{\!2}y^2 + o(y^2),
\label{eq:grating}
\end{equation}
where $p$ is the grating line density, $D$ the detector width, $f$ the camera focal length, 
and $\beta_0<0$ the diffraction angle at the central wavelength $\lambda_0$. The grating is used at its (+1) diffraction order. Substituting 
Eq.~\eqref{eq:grating} into Eq.~\eqref{eq:sag_taylor} gives the intermediate form:
\begin{align}
\Delta z = \frac{2n}{\mathrm{NA}^2}\Bigg[
&\frac{d(\delta_\lambda W_{020})}{d\lambda}\bigg|_{\lambda_0}
\!\left\{\frac{D\cos\beta_0}{2pf}y - \frac{D^2\sin\beta_0}{8pf^2}y^2 
+ o(y^2)\right\} + \frac{1}{2}\frac{d^2(\delta_\lambda W_{020})}{d\lambda^2}\bigg|_{\lambda_0}
\!\left\{\frac{D\cos\beta_0}{2pf}y + o(y)\right\}^{\!2} \nonumber\\
&+ W^{\rm cam}_{220m}\left(y - a^{\rm cam}_{220m}\right)^2 
+ \left(W^{\rm col}_{220m} + W^{\rm cam}_{220m}\right)x^2 \nonumber\\
&+ W^{\rm col}_{220m}\!\left({a^{\rm col}_{220m}}^{\!2} + b^{\rm col}_{220m}\right) 
+ W^{\rm cam}_{220m}b^{\rm cam}_{220m}
\Bigg],
\label{eq:sag_intermediate}
\end{align}
Collecting terms by order in $x$ and $y$ then yields the final expression for the ideal 
detector sag:
\begin{align}
\Delta z = \frac{2n}{\mathrm{NA}^2}\Bigg[
&\underbrace{\left(
W^{\rm cam}_{220m} 
- \frac{D^2\sin\beta_0}{8pf^2}\frac{d(\delta_\lambda W_{020})}{d\lambda}\bigg|_{\lambda_0} 
+ \frac{1}{8}\!\left(\frac{D\cos\beta_0}{pf}\right)^{\!2}
\frac{d^2(\delta_\lambda W_{020})}{d\lambda^2}\bigg|_{\lambda_0}
\right)}_{\text{spectral curvature}}\,y^2 +\; \underbrace{\Bigg(W^{\rm col}_{220m} + W^{\rm cam}_{220m}\Bigg)}_{\text{spatial curvature}}\,x^2 \nonumber\\
&+\; \underbrace{\left(
\frac{D\cos\beta_0}{pf}\frac{d(\delta_\lambda W_{020})}{d\lambda}\bigg|_{\lambda_0} 
- 2W^{\rm cam}_{220m}a^{\rm cam}_{220m}
\right)}_{\text{detector tilt}}\,y \nonumber\\
&+\; \underbrace{
W^{\rm col}_{220m}\!\left({a^{\rm col}_{220m}}^{\!2} + b^{\rm col}_{220m}\right) 
+ W^{\rm cam}_{220m}\!\left(b^{\rm cam}_{220m} + {a^{\rm cam}_{220m}}^{\!2}\right)
}_{\text{defocus offset}}
\Bigg],
\label{eq:sag_final}
\end{align}
where $x$ and $y$ are the normalized coordinates along the spatial and spectral axes of the detector, respectively.

\subsubsection*{Physical interpretation}

Four distinct contributions emerge from Eq.~\eqref{eq:sag_final}, each with a clear physical origin:
\begin{itemize}
    \item \textbf{Spatial curvature} (term in $x^2$):  Corresponds to the sum of the collimator and camera field curvatures only.

    \item \textbf{Spectral curvature} (term in $y^2$): Composed of the camera field curvature on which is added the quadratic component of axial chromatism and its mapping on the detector. Because these corrections differ from the spatial curvature, \textbf{the ideal detector shape is generally a toroid, not a sphere}.

    It is worth noting the physical significance of the $-\frac{D^2\sin\beta_0}{8pf^2}\frac{d(\delta_\lambda W_{020})}{d\lambda}\big|_{\lambda_0}$ term in the spectral curvature coefficient of Eq.~\eqref{eq:sag_final}. Its negative sign originates from the negative quadratic coefficient in the grating dispersion relation, Eq.~\eqref{eq:grating}, which reflects the geometric compression of wavelength intervals on the image surface as diffraction angle increases. Consequently, contrary to what one might expect, this contribution couples the nonlinearity of the dispersion not to the second derivative of the axial chromatism, but to its first derivative. The sign of the resulting correction therefore depends on the axial chromatism of the camera: for a standard undercorrected system ($d(\delta_\lambda W_{020})/d\lambda < 0$, red focusing at larger z), this term is negative and \textbf{adds} to the spectral curvature, compounding the effect of the camera field curvature $W^{\text{cam}}_{220m}$, which is assumed to be negative due to the Petzval curvature of the camera. Conversely, an overcorrected camera ($d(\delta_\lambda W_{020})/d\lambda > 0$) yields a partial cancellation. This represents an interesting degree of freedom in instrument design: by controlling the sign and magnitude of the axial chromatism, it is in principle possible to partially compensate the spectral field curvature through the grating dispersion nonlinearity.

    \item \textbf{Global tilt} (term in $y$): Arises from the combined effect of off-axis camera elements and the linear component of axial chromatism. It must be accommodated by tilting the detector.

    \item \textbf{Defocus offset} (constant term): Present only when optical components are used off-axis; correctable by a simple axial shift of the detector.
\end{itemize}

It should be noted that this analysis assumes a \textbf{thin, flat input slit} and is limited to \textbf{fourth-order} aberrations. We also neglected the \textbf{spectral anamorphism} \big($\frac{\mathrm{cos}(\alpha)}{\mathrm{cos}(\beta(\lambda))}$\big) induced by the grating, as well as \textbf{smile and keystone distortion}. In practice, IFS spectrographs use a staircase-shaped pseudo-slit, and extending the expansion to the sixth order would introduce sixth order field curvature along with terms proportional to $y^3$ and $y^4$ through the Taylor expansion of axial chromatism. As shown in Sec.~\ref{sec:curvedDesign}, these residuals are small in our design but not entirely negligible.\\

The key conclusion is that the ideal detector surface for a spectrograph is a \textbf{tilted toroid}: curved differently along the spatial and spectral directions, and globally tilted due to the interplay of off-axis optics and axial chromatism. This provides the theoretical motivation for the curved-detector designs explored in the following section.

\subsection{Proposed optical design}
\label{sec:curvedDesign}

\begin{figure} [H]
\begin{center}
\begin{tabular}{c} 
\includegraphics[height=5.8cm]{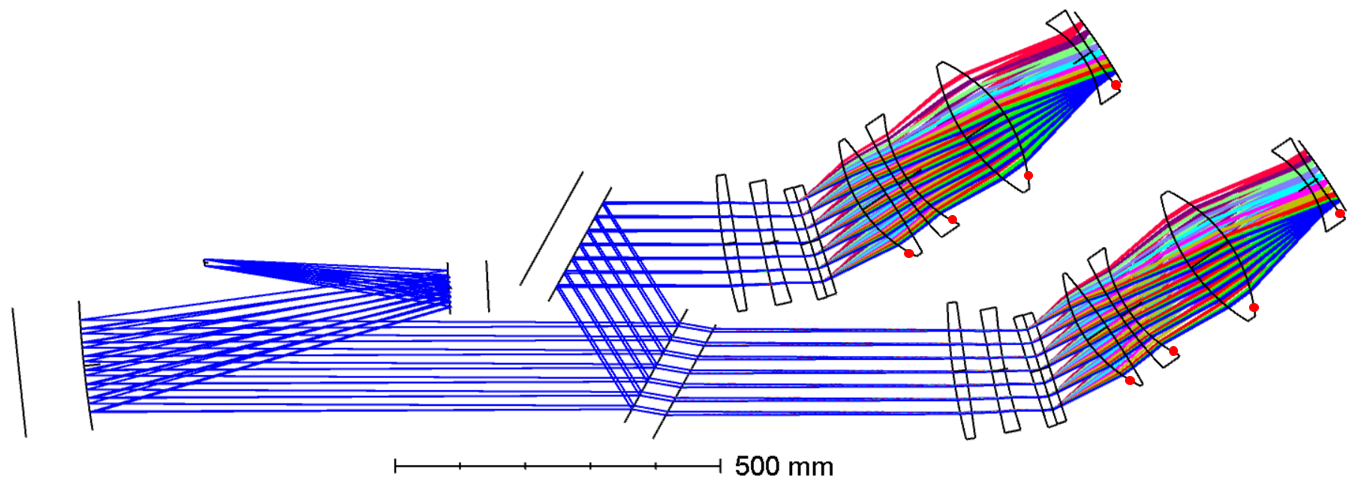}
\end{tabular}
\end{center}
\caption[SC] 
{ \label{fig:SC} 
Spectrograph layout for the design using curved detectors. The red dots correspond to aspherics.}
\end{figure}

A central question driving these designs is whether curved detectors offer tangible benefits, and whether those benefits justify the additional risk in terms of manufacturing, defects, and cost. Building on the previous design and allowing the detector to adopt a toroidal shape during optimization yields the design shown in Fig.~\ref{fig:SC}.

This second design consequently retains much of the same overall architecture as its predecessor. Notably, two lenses have been removed from the cameras in both arms. While one might expect the field lens to be eliminated in this configuration, the cryostat window enclosing the detector always necessitates a lens at that position. Beyond sealing the cryostat, this lens plays an active optical role: it helps correct field aberrations in the camera and, more critically, compensates part of the spectrograph's field curvature — thereby reducing the curvatures required on both the slit and the detector and limiting the associated manufacturing risks. The optimization also reveals that this design naturally converges toward a tilted toroidal rather than a tilted spherical detector geometry, in agreement with the theoretical framework established in the previous section. That said, a toroidal curvature introduces even greater fabrication risk, as this detector shape has been much less explored than spherical curvature \cite{TorCurved}.

To mitigate this risk, we explored partial compensation of the field curvature along the spatial axis through other elements of the spectrograph. The most straightforward approach is to introduce curvature on the input slit, which gives rise to two additional curving options summarized in Tab.~\ref{tab:CurvCases}.

\begin{table}[H]
\caption{Curving cases for the detector and the input slit.} 
\label{tab:CurvCases}
\begin{center}       
\begin{tabular}{|c|c|c|c|c|c|} 
\hline
\rule[-1ex]{0pt}{3.5ex}  Detector shape & Arm & Spatial curvature & Spectral curvature & Slit shape & Slit curvature  \\
\hline
\rule[-1ex]{0pt}{3.5ex}  \multirow{2}{*}{Toroidal} & Blue & -2677 mm & -846 mm & \multirow{2}{*}{Flat} & \multirow{2}{*}{None}  \\
\cline{2-4}
\rule[-1ex]{0pt}{3.5ex}  & Red & -2937 mm & -864 mm &  &   \\
\hline
\rule[-1ex]{0pt}{3.5ex}  \multirow{2}{*}{Cylindrical} & Blue & None & -881 mm & \multirow{2}{*}{Concave} & \multirow{2}{*}{3252 mm}  \\
\cline{2-4}
\rule[-1ex]{0pt}{3.5ex}  & Red & None & -875 mm &  &   \\
\hline
\rule[-1ex]{0pt}{3.5ex}  \multirow{2}{*}{Spherical} & Blue & N/A & N/A & \multirow{2}{*}{Convex} & \multirow{2}{*}{N/A}  \\
\cline{2-4}
\rule[-1ex]{0pt}{3.5ex}  & Red & N/A & N/A &  &   \\
\hline
\end{tabular}
\end{center}
\end{table}

The first configuration compensates the spatial field curvature via the slit and the spectral field curvature via the detector, resulting in a concave slit paired with a cylindrically curved detector. This option is especially attractive because curving the detector in only one direction is considerably simpler to achieve than imposing a spherical or toroidal shape \cite{Olaf2}. The second configuration calls for a spherical detector combined with a curved slit. This option is the least compelling: it carries manufacturing risks comparable to the previous case on the slicer side (see Sec.~\ref{sec:risk}), while imposing even greater demands on the detector, and it delivers inferior image quality relative to the other two configurations.

In terms of optical performance, the design with cylindrical detectors achieves lower image quality than the toroidal case due to how the blue and red arms' spatial curvatures are compensated. With toroidal detectors they are compensated individually, whereas with the cylindrical detectors, the slit curvature tries to compensate the two together. Both perform slightly below the flat case but remain highly competitive, with median \textbf{as-designed} spot sizes of 0.060 and 0.058 arcsec FWHM in the blue and red arms respectively, for the cylindrical case. This is achieved with moderately stronger aspheric surfaces, reaching a maximum departure of 0.7 mm from the best-fit sphere. The slight degradation relative to the flat-detector design mainly stems from the staircase profile of the slit \cite{slicer}; a thin slit would yield improved spot sizes.\\
As a result, we pursued the designs focusing on the toroidal and cylindrical curvature cases, both offering very similar performance, with a slight advantage for the toroidal case at the expense of a higher manufacturing risk.

\begin{figure} [H]
\begin{center}
\begin{tabular}{c} 
\includegraphics[height=6.5cm]{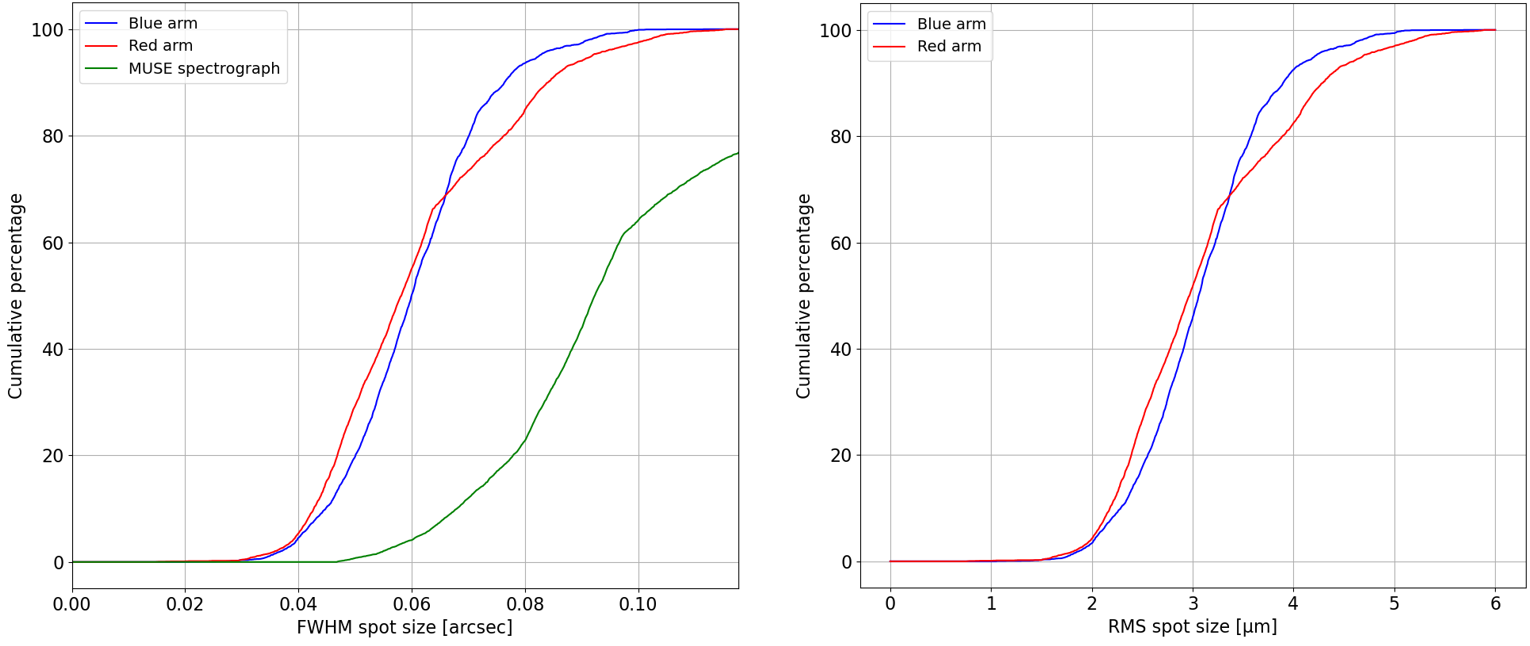}
\end{tabular}
\end{center}
\caption[SF] 
{ \label{fig:IQ_curved} 
Cumulative distribution of the FWHM spot size in the blue and red arms for the \textbf{as-designed (nominal)} spectrograph with cylindrically curved detectors, projected on sky (left) and on the detector (right).}
\end{figure} 

\subsection{Residuals between the curved detectors and the real image surface}

The theoretical analysis of Sec.~\ref{sec:theo} predicts that the ideal detector shape is dominated by low-order terms with smaller contributions from higher-order aberrations. To verify this and quantify how well the toroidal and cylindrical shapes approximate the true image surface, we computed the actual image surface directly from the optical model. This was done by freezing all other variables in the spectrograph, replacing the image surface description with an XY polynomial up to 6th order, and re-optimizing the detector shape alone. The difference between this freeform reference surface and the toroidal or cylindrical shapes obtained in the designs of Sec.~\ref{sec:curvedDesign} then gives a direct measure of the error introduced by constraining the detector to a simpler geometry.

\begin{figure} [H]
\begin{center}
\begin{tabular}{c} 
\includegraphics[height=6.2cm]{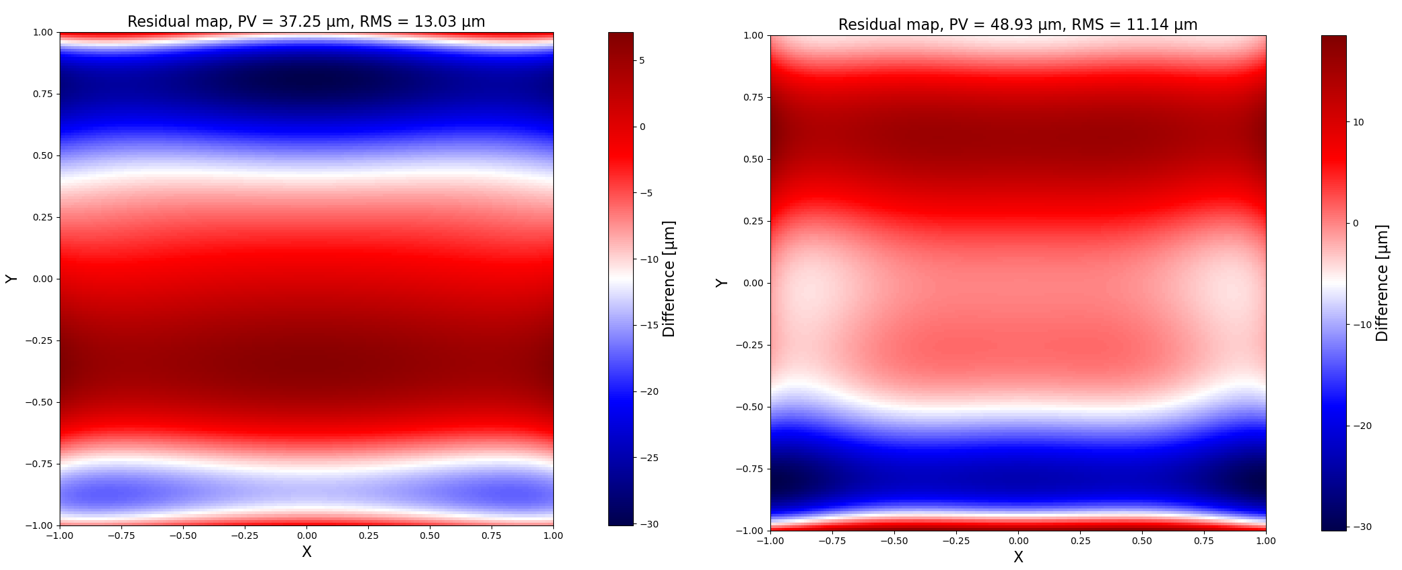}
\end{tabular}
\end{center}
\caption[SF] 
{ \label{fig:TOR_diff} 
Height differences between the best toroidal image surface and the real image surface shape in the blue (left) and red (right) arms.}
\end{figure}

\begin{figure} [H]
\begin{center}
\begin{tabular}{c} 
\includegraphics[height=6.2cm]{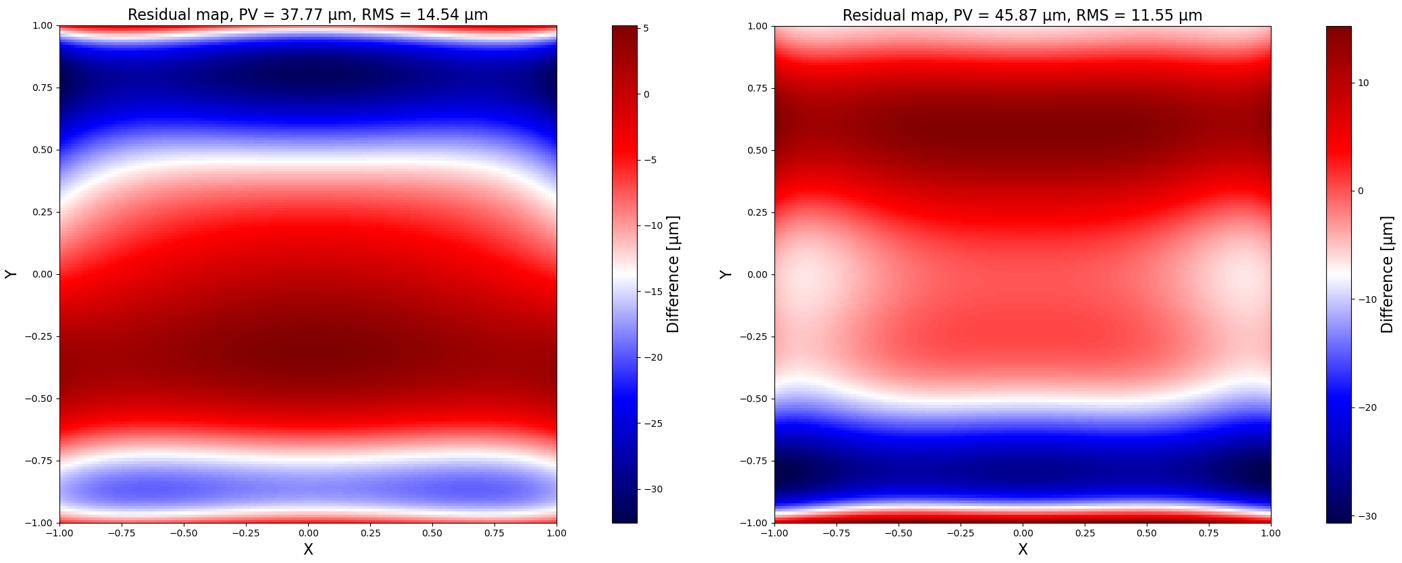}
\end{tabular}
\end{center}
\caption[SF] 
{ \label{fig:CYL_diff} 
Height differences between the best cylindrical image surface and the real image surface shape in the blue (left) and red (right) arms.}
\end{figure}

As shown in Figs.~\ref{fig:TOR_diff} and~\ref{fig:CYL_diff}, the actual image surface is indeed overwhelmingly dominated by quadratic and lower-order terms, in full agreement with the theoretical prediction. The residuals from higher-order terms are at most 59\,\textmu m peak-to-valley and 14.5\,\textmu m RMS, which is of the same order of magnitude as the detector's flatness error, assumed to be about $\pm10 \mu m$ PV. As expected, these residuals have a negligible impact on image quality: the differences in FWHM spot size between the cylindrical or toroidal designs and a freeform detector are insignificant.

The residuals are dominated by terms in $y^3$, $y^4$, $y^5$, and $y^6$. This is expected: as discussed in Sec.~\ref{sec:theo}, these powers of $y$ arise from two sources. The first is the off-axis use of optical components in the camera, which generates nodal field-curvature contributions along the spectral direction. The second and most important one is axial chromatism, which, given the width of our spectral ranges, cannot be perfectly captured by a second-order expansion in wavelength, leaving residual higher-order terms along $y$.

Taken together, these results confirm that toroidal and cylindrical detector shapes are excellent approximations to the true image surface for this design, and that the gains from adopting a more complex freeform detector would be marginal.

\subsection{Risk analysis}
\label{sec:risk}

Regarding the additional risk for the image slicer design mentioned earlier,  it is worth recalling the nature of the slit at its output. It consists of slitlets — small individual slits corresponding to the sliced fields — whose shape and position are controlled independently rather than collectively. In our case, the spectrograph requires its input slit to be curved and telecentric. At the slitlet level, this translates into a requirement for each slitlet to follow the local field curvature at its position. More precisely, each slitlet must produce the appropriate nodal aberrations to ensure that a significant portion of it does not fall out of focus. At this stage, this aspect has not yet been assessed for WST, but it will be studied in the near future and will help determine whether the use of curved detectors in these spectrographs should be supported. Previous work on the design of the HARMONI image slicer for the ELT has shown that, in their specific case, a theoretical solution exists but remains prohibitively expensive to implement. \cite{Hslicer}

From a manufacturing risk perspective, comparable designs with larger pupils (200 mm versus 150 mm) were reviewed by industrial partners, whose assessment was that, aside from the detectors themselves, these designs raise very little concern. As summarized in Tab.~\ref{tab:RISK}, the risk levels across material availability, optical coatings, and MAIT are negligible or low overall, with the only unresolved item being the curved detector.

\begin{table}[ht]
\caption{Preliminary internal risk analysis provided by the manufacturers.} 
\label{tab:RISK}
\begin{center}       
\begin{tabular}{|c|c|c|c|} 
\hline
\rule[-1ex]{0pt}{3.5ex}  Risk & Description & Design with a flat detector & Design with a curved detector  \\
\hline
\rule[-1ex]{0pt}{3.5ex}  A & Material Availability (technical) & None & None   \\
\hline
\rule[-1ex]{0pt}{3.5ex}  B & Material Availability (production) & Low & Low  \\
\hline
\rule[-1ex]{0pt}{3.5ex}  C & Component manufacturing & None & None \\
\hline
\rule[-1ex]{0pt}{3.5ex}  D & Optical coatings & None & None \\
\hline
\rule[-1ex]{0pt}{3.5ex}  E & Integration and tests & None & None \\
\hline
\rule[-1ex]{0pt}{3.5ex}  F & Sensor & None & Unknown \\
\hline
\end{tabular}
\end{center}
\end{table}

\section{CONCLUSION}

This paper describes the current state of progress on the optical design for the WST integral-field spectrograph, with an emphasis on the design of the spectrograph subsystem and the potential gains from curved detector technology.\\
The baseline flat-detector design already delivers strong performance, using a compact, mirror-based collimator and a camera with a limited number of optical surfaces. Building on this, we developed a theoretical framework for the ideal detector shape in a spectrograph, showing analytically that it takes the form of a tilted toroid whose curvatures along the spatial and spectral axes arise from distinct physical contributions: collimator and camera field curvature, axial chromatism, and the nonlinearity of grating dispersion.\\
Of the three curved-detector configurations explored, the cylindrical detector paired with a curved input slit emerges as the most practically viable. It achieves near-equivalent optical performance to the toroidal case while imposing significantly lower fabrication demands, and enables the removal of two lenses per spectrograph arm, translating to 768 fewer optical elements at full instrument scale. This represents a meaningful reduction in both cost --- a savings of approximately €7M --- and optical surfaces contributing to throughput losses, without any fundamental change in the spectrograph architecture.\\
The principal outstanding risk lies in the curved detector itself, and its interaction with the image slicer design. Curved slitlets at the slicer output will need to satisfy local field curvature requirements that have not yet been assessed for WST, and will be the subject of future study. The toroidal case, while currently considered too risky by the consortium, should not be forgotten: the required curvatures are modest compared to what has been demonstrated, and it would avoid the need for a more complex slicer design entirely. As fabrication techniques mature, it may become the preferred solution.\\
Taken together, these results establish a technically credible and cost-effective baseline for the WST IFS spectrographs, and provide a transferable analytical framework that can guide the design of future large-scale integral field instruments.

\acknowledgments 
 
This project has received funding from the European Union’s Horizon Europe research and innovation programme under grant agreement No 101183153.

\bibliography{report} 
\bibliographystyle{spiebib} 

\end{document}